\documentclass[aps,prd,groupedaddress,showpacs,tighten,floats,twocolumn,nofootinbib]{revtex4}
\usepackage{graphicx}
\usepackage{latexsym}
\def\beq{\begin{equation}}
\def\eeq{\end{equation}}
\def\bey{\begin{eqnarray}}
\def\eey{\end{eqnarray}}

\def\sigv{\langle\sigma v\rangle}

\def\lsim{\mathrel{\raise.3ex\hbox{$<$\kern-.75em\lower1ex\hbox{$\sim$}}}}
\def\gsim{\mathrel{\raise.3ex\hbox{$>$\kern-.75em\lower1ex\hbox{$\sim$}}}}

\begin{document}

\title{Constraining Supersymmetric Dark Matter With Synchrotron Measurements}  
\author{Dan Hooper}
\address{Fermi National Accelerator Laboratory, Theoretical Astrophysics, Batavia, IL  60510}

\date{\today}

\begin{abstract}

The annihilations of neutralino dark matter (or other dark matter candidate) generate, among other Standard Model states, electrons and positrons. These particles emit synchrotron photons as a result of their interaction with the Galactic Magnetic Field. In this letter, we use the measurements of the WMAP satellite to constrain the intensity of this synchrotron emission and, in turn, the annihilation cross section of the lightest neutralino. We find this constraint to be more stringent than that provided by any other current indirect detection channel. In particular, the neutralino annihilation cross section must be less than $\approx 3\times 10^{-26}$cm$^3$/s ($1\times 10^{25}$cm$^3$/s) for 100 GeV (500 GeV) neutralinos distributed with an NFW halo profile. For the conservative case of an entirely flat dark matter distribution within the inner 8 kiloparsecs of the Milky Way, the constraint is approximately a factor of 30 less stringent. Even in this conservative case, synchrotron measurements strongly constrain, for example, the possibility of wino or higgsino neutralino dark matter produced non-thermally in the early universe.

\end{abstract}
\pacs{95.35.+d;95.30.Cq,95.55.Ka; FERMILAB-PUB-08-019-A}
\maketitle


If dark matter consists of particles with a weak-scale mass and couplings, then their annihilations are expected to produce a variety of potentially observable particles, including gamma rays~\cite{gamma}, neutrinos~\cite{neutrinos}, positrons~\cite{positron}, antiprotons~\cite{antiproton}, antideuterons~\cite{antideu}, X-rays~\cite{Bergstrom:2006ny} and synchrotron radiation~\cite{syn,darkhaze1}. The synchrotron emission resulting from dark matter annihilations naturally falls in the frequency range studied by cosmic microwave background (CMB) missions, such as the Wilkinson Microwave Anisotropy Probe (WMAP)~\cite{spergel}. Data from WMAP and other CMB experiments can, therefore, be used to potentially constrain or detect the presence of dark matter annihilations in our galaxy. 

It has been previously argued that microwave emission observed from the inner Milky Way by WMAP (the ``WMAP Haze'') is likely the product of dark matter annihilations~\cite{darkhaze,darkhaze1} (see also Ref.~\cite{doblerfink}). In this letter, we do not take this conclusion for granted, but instead simply use the WMAP data to place an upper limit on the rate of dark matter annihilation taking place in the inner kiloparsecs of the Milky Way. In particular, we focus on supersymmetric neutralinos as our dark matter candidate. As we will show, the properties of such particles can be meaningfully constrained by the degree of synchrotron emission observed by WMAP.

Assuming that neutralinos constitute a large fraction of the galactic dark matter, the rate of neutralino annihilations taking place within a distance, $R_{\rm max}$, from the center of the Milky Way is given by:
\begin{equation}
R_{\chi} = 2 \pi\int^{r_{\rm max}}_0 \frac{\rho^2(r)\, \sigv}{m^2_{\chi}} r^2 dr,
\end{equation}
where $\rho(r)$ is the density of dark matter at a distance, $r$, from the Galactic Center, $\sigv$ is the thermally averaged neutralino annihilation cross section (multiplied by the relative velocity) and $m_{\chi}$ is the neutralino's mass. Depending on the details of the supersymmetric model, neutralino annihilations lead to a variety of final states, dominated by a combination of heavy fermions ($b\bar{b}$, $t\bar{t}$, $\tau^+ \tau^-$) and gauge and/or Higgs bosons~\cite{jungman}. When produced, these particles fragment and decay, leading to a combination of photons, electrons, protons, neutrinos and their antiparticles. The electrons and positrons which are produced then proceed to travel under the influence of the Galactic Magnetic Field, losing energy via inverse Compton and synchrotron processes. The resulting flux of synchrotron emission is given by:
\begin{equation}
F_{\rm syn} = \frac{F_e\, F_{\rm cont}\,R_{\chi}}{m_{\chi}} \frac{U_{B}}{U_{B}+U_{\rm rad}},
\label{synpower}
\end{equation}
where $F_e$ denotes the fraction of the annihilation power that goes into electrons and positrons and $F_{\rm cont}$ is the average fraction of the electron's energy which is radiated (via synchrotron or inverse Compton) before it leaves the region of interest. In the case of WMAP's observation of the inner Milky Way, this quantity is expected to be near unity.

\begin{figure}

\resizebox{7.5cm}{!}{\includegraphics{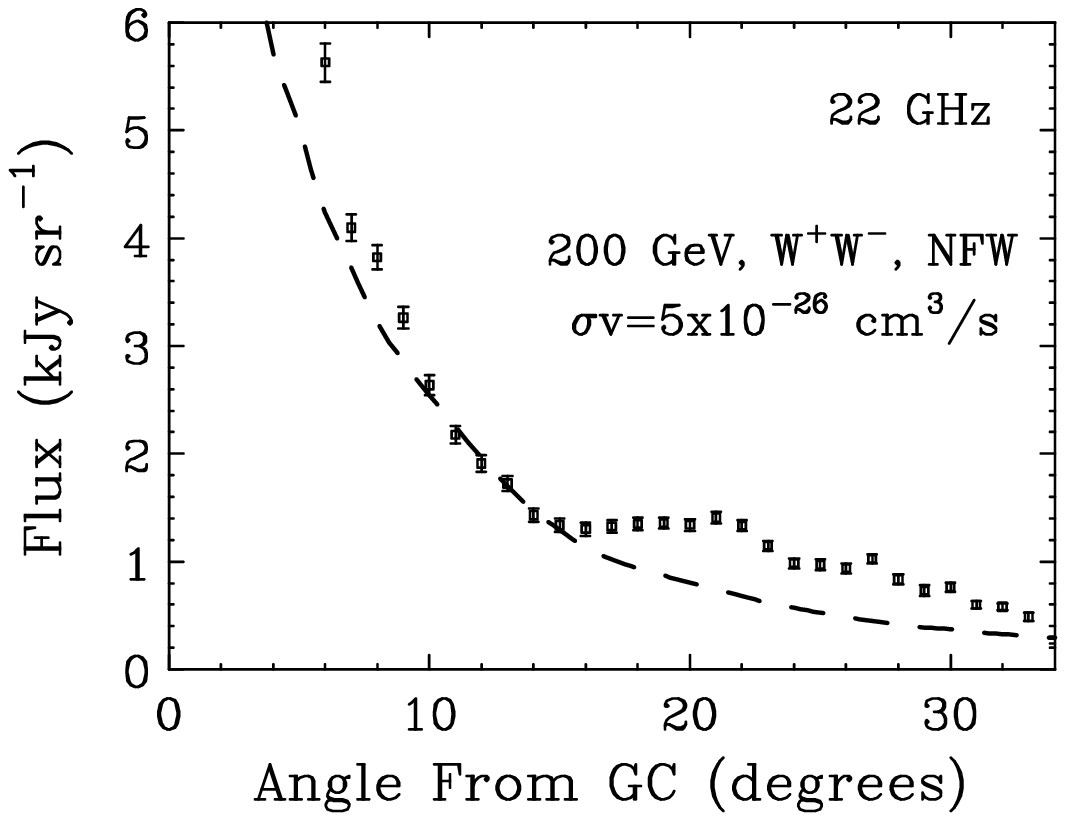}}
\\ 
\vspace{0.5cm}
\resizebox{7.5cm}{!}{\includegraphics{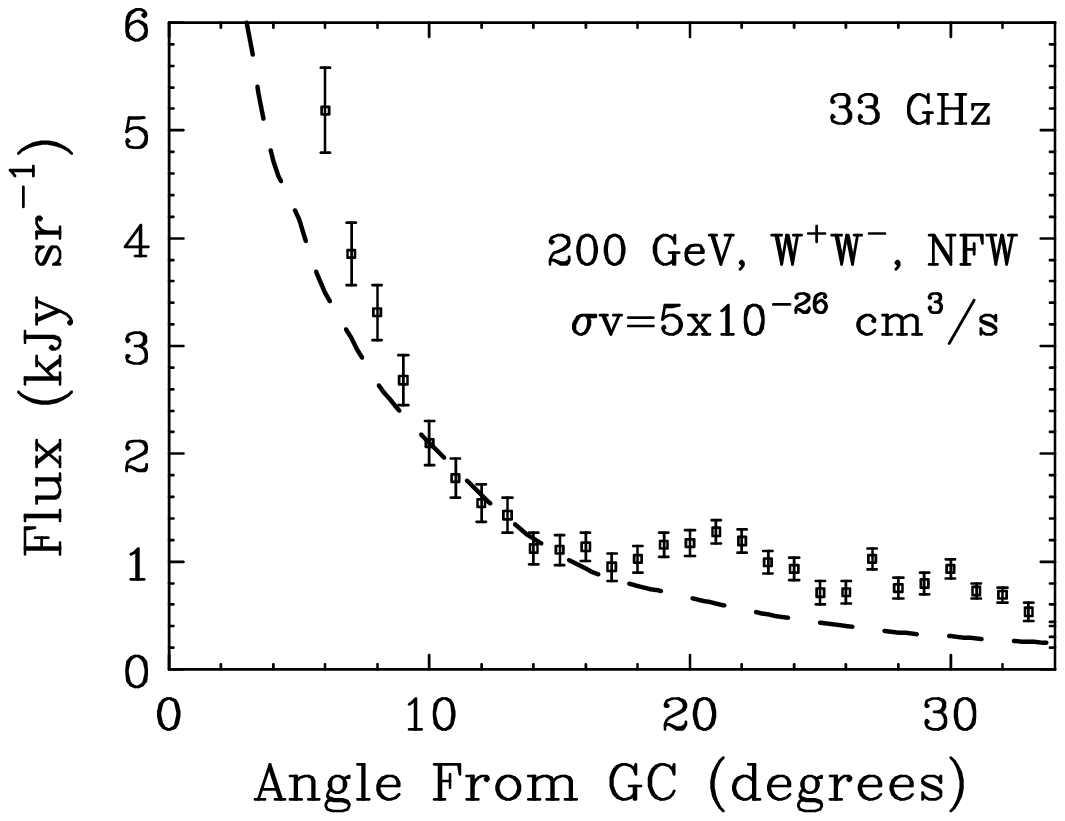}}
\caption{The specific intensity (in kilo-Janskys per steradian) observed by WMAP in its 22 and 33 GHz bands, as a function of the angle from the Galactic Center. In each frame, the dashed line denotes the flux of synchrotron emission from the annihilation products of a 200 GeV neutralino annihilating to $W^+ W^-$ with an annihilation cross section of $\sigma v = 5 \times 10^{-26}$ cm$^3$/s and distributed with an NFW halo profile. We have used $U_B/(U_B+U_{\rm rad}) = 0.26$ and the diffusion parameters described in the text.
}
\label{start}
\end{figure}

$U_{B}$ and $U_{\rm rad}$ are the energy densities of magnetic fields and radiation (starlight, emission from dust, and the CMB) in the inner Galaxy, respectively. Their role in Eq.~\ref{synpower} is to account for the fraction of the electrons' energy which is emitted as synchrotron, as opposed to inverse Compton scattering. These two processes yield similar energy loss rates. For example, in the local region of our galaxy, $B_{\rm rms} \sim 3 \, \mu$G and $U_{\rm rad} \approx 0.9$ eV/cm$^3$ (0.3 and 0.6 eV/cm$^3$ from the cosmic microwave background and starlight, respectively), leading to $U_{B}/(U_{B}-U_{\rm rad}) \approx 0.18$. $U_B$ and $U_{\rm rad}$ are larger in the inner Galaxy, but the ratio is not expected to change dramatically. At 2-3 kiloparsecs from the Galactic Center, for example, reasonable estimates of $B_{\rm rms} \sim10 \, \mu$G and $U_{\rm rad} \sim 5$ eV/cm$^3$~\cite{isrf} yield  $U_{B}/(U_{B}-U_{\rm rad}) \approx 0.26$.

The angular distribution of synchrotron emission produced through neutralino annihilations depends on both the spatial distribution of dark matter and on the propagation of electrons in the halo ({\it ie.}, the geometry of the Galactic Magnetic Field). Following Ref.~\cite{darkhaze}, we start by considering an Navarro-Frenk-White (NFW)~\cite{nfw} halo profile as a benchmark, and adopt a diffusion constant of $K(E_e) \approx 10^{28} \, (E_{e} / 1 \, \rm{GeV})^{0.33} \,\rm{cm}^2 \, \rm{s}^{-1}$ and an average electron energy loss time of $b(E_e) = 5 \times 10^{-16} \, ({E_e} / 1 \, \rm{GeV})^2 \,\, \rm{s}^{-1}$. For calculating the synchrotron spectrum, we use a 10 $\mu$G magnetic field. We arrive at the results shown in Fig.~\ref{start}. Here, we have considered a 200 GeV neutralino which annihilates to $W^+ W^-$ with a cross section of $\sigma v = 5\times 10^{-26}$ cm$^3$/s. This cross section was chosen because it leads to a synchrotron flux that saturates the WMAP observations over angles of 10$^{\circ}$ to 15$^{\circ}$ from the Galactic Center. If the cross section were significantly larger, the model would predict a synchrotron flux inconsistent with WMAP.

Results are shown in Fig.~\ref{start} for two of the five WMAP frequency bands, 22 and 33 GHz. The error bars in the other (higher) frequency bands are somewhat larger~\cite{darkhaze,doblerfink} and thus are less useful in placing constraints on the contribution from dark matter annihilations.

\begin{figure}
\hspace{-0.6cm}
\resizebox{9.0cm}{!}{\includegraphics{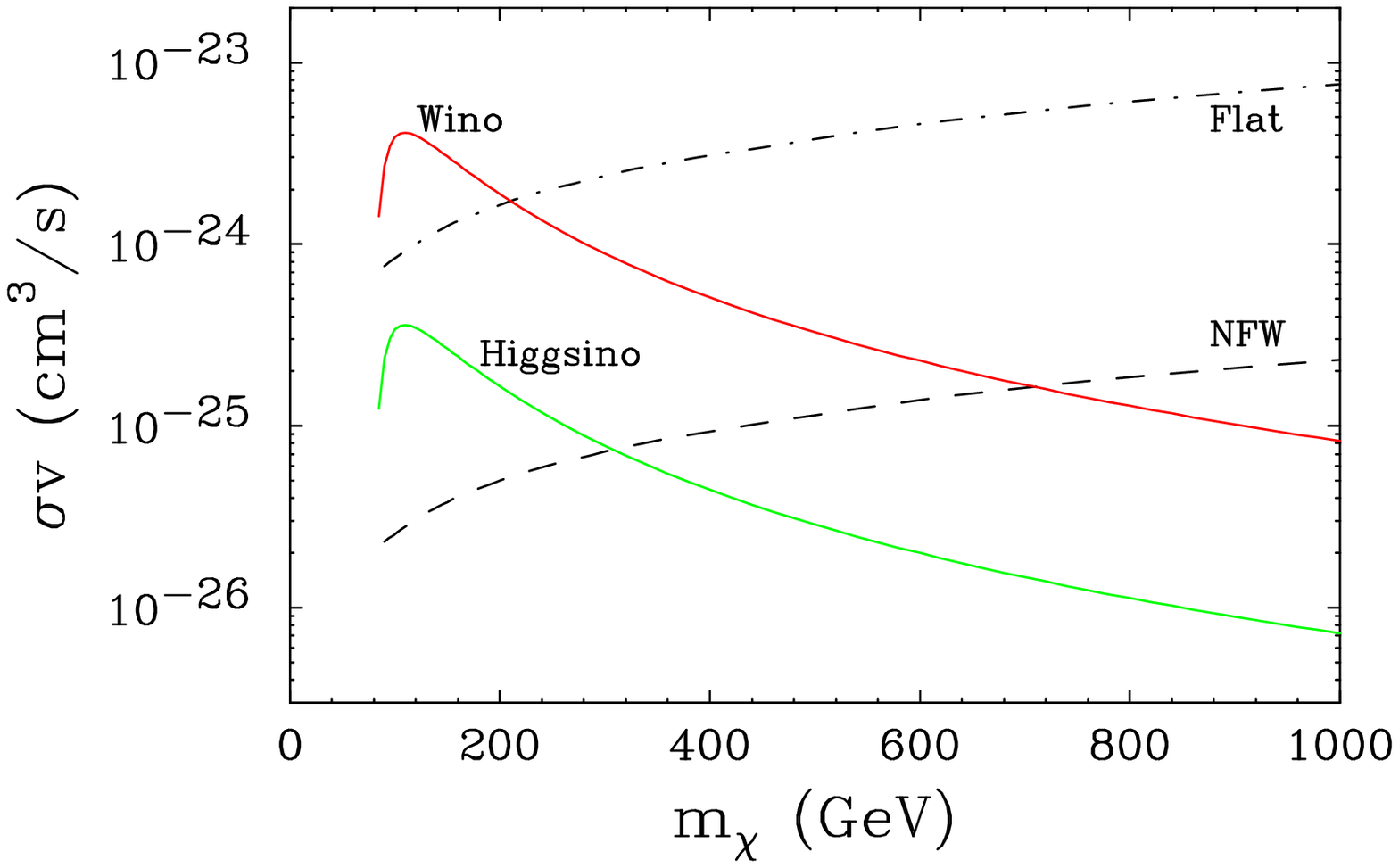}}\\
\hspace{-0.6cm}
\vspace{0.5cm}
\resizebox{9.0cm}{!}{\includegraphics{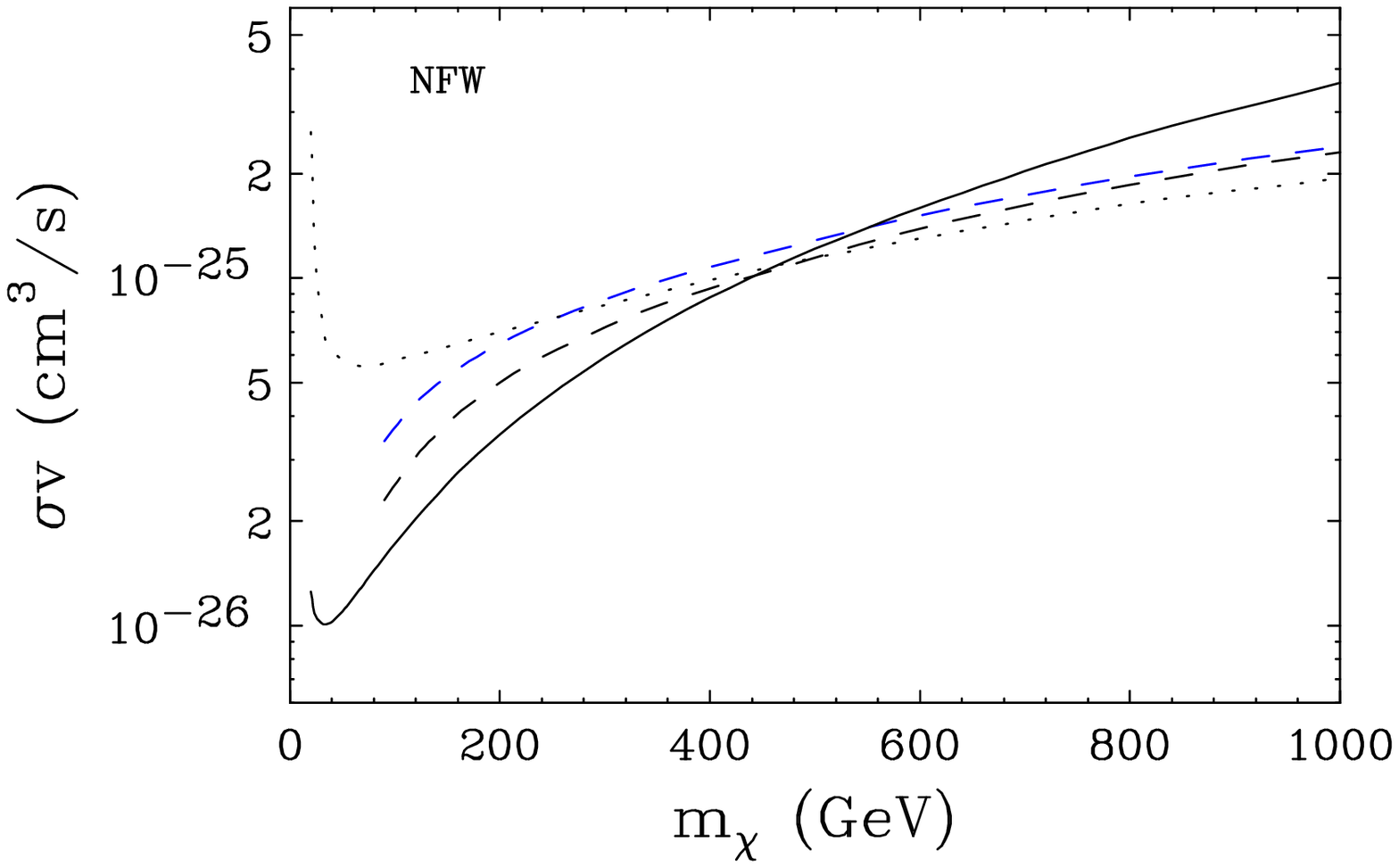}}
\caption{Top: The upper limit on the neutralino annihilation cross section from the synchrotron constraint as a function of mass, for the case of an NFW halo profile (dashed) and a flat (homogeneous) distribution of dark matter within the solar circle (dot-dashed). These limits were arrived at considering neutralinos which annihilate largely to $W^+ W^-$ (as is the case for wino or higgsino-like neutralinos), $U_B/(U_B+U_{\rm rad}) = 0.26$ and the diffusion parameters described in the text. Shown for comparison are the annihilation cross sections for a pure-wino (red solid) and a pure-higgsino (green solid). Bottom: The upper limit found with an NFW profile, and for several dominant annihilation modes, $b\bar{b}$ (dotted), $ZZ$ (blue dashed), $W^+W^-$ (black dashed) and $\tau^+ \tau^-$ (solid).}
\label{limit}
\end{figure}

In the upper frame of Fig.~\ref{limit}, we show as a dashed line the upper limit from synchrotron emission in the inner Galaxy on the neutralino annihilation cross section as a function of mass for the case of annihilations to $W^+ W^-$. In the lower frame of Fig.~\ref{limit}, we show the constraint for other common neutralino annihilation modes. The constraints shown here are quite stringent, especially in the case of light neutralinos. The strength of this constraint depends strongly, however, on the way in which the dark matter is distributed in the Inner Galaxy. 

As the gravitational potential in inner kiloparsecs of the Milky Way is dominated by baryons rather than dark matter, it is difficult to place significant observational constraints on the distribution of dark matter in this region. Although numerical simulations indicate that high density cusps (such as that found in the NFW profile) are expected to be present, we do not take this for granted here. Observations of the rotation curves of our Galaxy do, however, constrain the total mass of dark matter inside of the solar circle (within $\approx 8$ kpc)~\cite{local}. As a highly conservative example, we will consider the scenario in which the dark matter inside of the solar circle is distributed homogeneously. With such a flat distribution, the annihilation rate is reduced considerably, leading to a synchrotron constraint a factor of $\sim 30$ less stringent compared to the NFW case. In the upper frame of Fig.~\ref{limit}, the dot-dashed line denotes the upper limit for the case of a flat dark matter distribution within the solar circle.

To be thermally produced in the early universe with an abundance consistent with the observed density of dark matter, a neutralino must annihilate  with a cross section of $\sigv \sim 3 \times 10^{-26}$ cm$^3$/s at the temperature of freeze-out (typically about 1/20 of the neutralino mass). The annihilation cross section of thermally produced neutralinos in the galactic halo ({\it ie.} in the low velocity limit) is, therefore, expected to be not much larger than $\sigv \sim 3 \times 10^{-26}$ cm$^3$/s, and possibly smaller. The limits shown in Fig.~\ref{limit} for the conservative case of a flat profile thus do not strongly constrain scenarios in which the dark matter is produced thermally. 

Neutralino dark matter could also be produced via non-thermal mechanisms, however.  For example, late-time decays of gravitinos, Q-balls or other such states could populate the universe with neutralino dark matter well after thermal freeze-out has occurred~\cite{latedecay}. Furthermore, as the thermal history of our universe has not been observationally confirmed back to the time of dark matter's chemical decoupling, one could also imagine a scenario in which neutralinos with a very large annihilation cross section were produced with the measured dark matter abundance due to a faster than expected expansion rate at freeze-out, or other non-standard cosmology~\cite{nonstandard}.

Neutralinos whose composition is dominantly wino or higgsino have particularly large annihilation cross sections. The lightest neutralino in the Anomaly Mediated Supersymmetry Breaking (AMSB) scenario, for example, is a nearly pure wino. Neutral winos annihilate very efficiently through the t-channel exchange of a nearly degenerate chargino. The cross section for the process, in the low velocity limit, is given by:
\begin{eqnarray}
\sigma v(\chi \chi \rightarrow W^+ W^-) &\approx& \frac{g^4 (m^2_{\chi}-m^2_{W})}{2 \pi m^2_{\chi}   (2 m^2_{\chi}-m^2_W)^2  } \\ \nonumber
&\sim& 1.7 \times 10^{-24} \, {\rm cm}^3/{\rm s} \,\times \bigg(\frac{200 \, {\rm GeV}}{m_{\chi}}\bigg)^2,
\end{eqnarray} 
which is much larger than the cross section required of a thermally produced dark matter candidate. Pure-higgsino neutralinos also annihilate very efficiently, resulting in both $W^+ W^-$ and $ZZ$ final states through the t-channel exchange of a chargino or neutralino, respectively. In the upper frame of Fig.~\ref{limit}, we compare the limits presented here to the predicted cross sections for a wino or higgsino neutralino. Even with the very conservative choice of a flat dark matter distribution, wino-like neutralino dark matter exceeds the synchrotron limit if $m_{\chi} \lsim 210$ GeV.

\begin{figure}
\vspace{0.5cm}
\hspace{-0.6cm}
\resizebox{9.0cm}{!}{\includegraphics{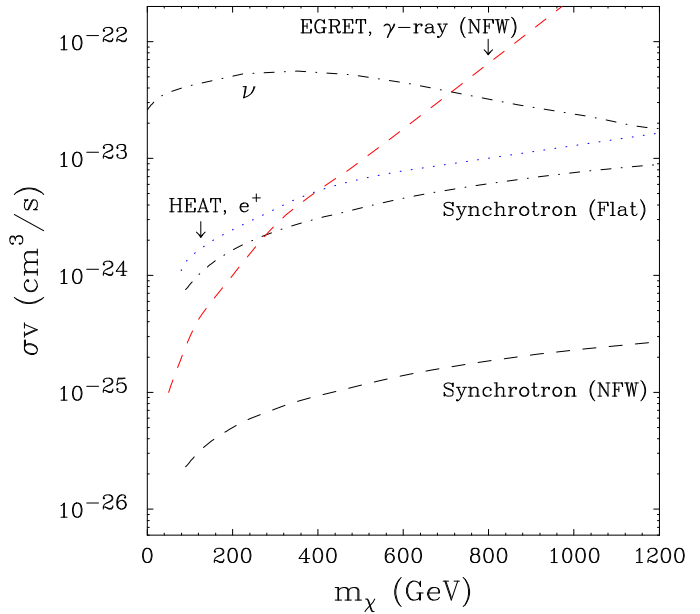}}
\caption{A comparison of the limits placed on the dark matter's annihilation cross section from several astrophysical channels. The black dashed and dot-dashed lines represent the synchrotron constraints (see Fig.~\ref{limit}) for the case of a NFW halo profile and the conservative case of a flat dark matter distribution, respectively. The dotted blue line represents the constraint which can be arrived at from measurements of the cosmic positron spectrum by the HEAT experiment~\cite{positron}. The red dashed line is the limit from the EGRET gamma-ray satellite for the case of an NFW halo profile~\cite{dingus}. The upper dot-dashed curve is the conservative limit from the diffuse neutrino flux, assuming dark matter annihilates only to neutrinos~\cite{beacom}.  With the exception of the neutrino constraint, each of these limits were arrived at considering neutralinos which annihilate largely to $W^+ W^-$ (as is the case for wino or higgsino-like neutralinos).}
\label{limitcompare}
\end{figure}

In Fig.~\ref{limitcompare}, we compare the constraint presented here with those obtained using other astrophysical observations. In particular, we show the upper limit on the dark matter annihilation cross section from the absence of gamma-rays observed from the Galactic Center by EGRET (for the case of an NFW halo profile)~\cite{dingus}, and the from observations of the cosmic positron spectrum~\cite{positron}. Each of these constraints are shown for the case of WIMPs annihilating to $W^+ W^-$. We also include, for comparison, the bound from the lack of observed diffuse neutrinos, as found in Ref.~\cite{beacom}, which corresponds to the conservative case in which WIMPs annihilate only to neutrinos. From this figure, we conclude that the synchrotron constraint calculated here is the more stringent than is found with any other channel.


To summarize, we have presented here a constraint on the annihilation cross section of neutralino dark matter derived from the observation of the inner Milky Way by WMAP. Dark matter annihilations produce relativistic electrons and positrons which generate synchrotron emission through their interactions with the Galactic Magnetic Field. By studying the intensity of radiation at synchrotron frequencies, an upper limit can be placed on the dark matter annihilation rate and corresponding annihilation cross section. We have compared the constraint presented here to that found from gamma-ray and positron observations, and find the limit from synchrotron emission to be the most stringent, even for the conservative case of a flat dark matter distribution within the solar circle. This constraint can be used to exclude dark matter candidates with large annihilation cross sections, such as wino or higgsino-like neutralinos produced through non-thermal mechanisms in the early universe.

\smallskip

We would like to thank Doug Finkbeiner and Greg Dobler for valuable discussions. This work has been supported by the US Department of Energy and by NASA grant NAG5-10842.

\end{document}